\documentclass[9pt,twocolumn,twoside]{opticajnl}
\journal{opticajournal} 

\setboolean{shortarticle}{true}

\usepackage{lineno}

\title{Ultralow-loss photonic integrated chips on 8-inch anomalous-dispersion Si$_3$N$_4$-SiO$_2$-Si Wafer}

\author[1,*]{Shuai Liu}
\author[2]{Matthew W. Puckett} 
\author[2]{Jianfeng Wu}
\author[1]{Abdulkarim Hariri}
\author[1]{Yuheng Zhang}
\author[1,*]{Zheshen Zhang}

\affil[1]{Department of Electrical Engineering and Computer Science, The University of Michigan, Ann Arbor, Michigan 48109, USA.}
\affil[2]{Honeywell International, Phoenix, Arizona, USA}
\affil[*]{shualiu@umich.edu; zszh@umich.edu}

\begin{abstract}

We report the fabrication of 8-inch crack-free, dispersion-engineered Si$_3$N$_4$-SiO$_2$-Si wafers fully compatible with industrial foundry silicon photonics fabrication lines. By combining these wafers with a developed amorphous silicon (a-Si) hardmask etching technique, we achieve ultra-low-loss Si$_3$N$_4$ photonic integrated circuits (PICs) with intrinsic quality factors exceeding $25 \times 10^6$ using electron beam lithography and $24 \times 10^6$  using standard ultraviolet stepper photolithography. Frequency-comb generation is demonstrated on these high-quality Si$_3$N$_4$ PICs, corroborating the designed anomalous dispersion. These results establish the feasibility of mass-manufacturing high-performance, dispersion-engineered Si$_3$N$_4$ PICs using standard foundry-grade processes, opening new pathways for applications in optical communications, nonlinear optics, and quantum optics.
\end{abstract}

\setboolean{displaycopyright}{false} 

\begin{document}

\maketitle

\section{Introduction}

Silicon photonics has emerged as a transformative platform for integrated optical technologies, offering compact, scalable, and energy-efficient solutions to meet the continuously increasing demands of high-data-rate telecommunications and data processing in modern data centers \cite{shi2020scaling,shi2022silicon}. Among various material platforms, silicon nitride (Si$_3$N$_4$) stands out as a leading candidate due to its unique combination of wide transparency window, low linear and nonlinear losses, moderate refractive index, strong Kerr nonlinearity, and complete compatibility with complementary metal-oxide-semiconductor (CMOS) fabrication processes \cite{ji2021methods}. In the past decades, Si$_3$N$_4$ PICs have unlocked a wide range of applications, including optical communication \cite{marin2017microresonator}, atomic clock \cite{newman2019architecture}, spectroscopy \cite{dutt2018chip}, quantum computing \cite{arrazola2021quantum, wu2020quantum}, and quantum sensing \cite{xia2023entanglement}. 

Low-confinement Si$_3$N$_4$ waveguides, typically with a thickness below 150 nm, represent a well-established and widely accessible platform in commercial foundries. Their large mode volumes  significantly reduce interaction with waveguide roughness, thus enabling record-low propagation loss down to 0.1 dB/m \cite{puckett2021422}. This exceptional performance makes them ideal for ultra-low-loss applications such as frequency metrology and narrow-linewidth lasers \cite{xiang20233d}, though they are hindered by large device footprints. Moderate-confinement Si$_3$N$_4$ waveguides, with thicknesses ranging from 150 nm to 400 nm, offer a balance between low propagation loss, mass production compatibility, and a compact footprint \cite{naraine2024moderate}. These platforms have recently become available in several industrial foundries, such as CUMEC \cite{zhang2024300}, Ligentec \cite{smith2023sin}, and AIM Photonics \cite{fahrenkopf2019aim}. However, many applications, such as broadband frequency-comb generation, require anomalous dispersion that usually necessitates Si$_3$N$_4$ thicknesses exceeding 600 nm \cite{ji2021methods, ye2023foundry}. Achieving such high-confinement Si$_3$N$_4$ waveguides presents significant fabrication challenges due to the high tensile stress in thick Si$_3$N$_4$ films (thickness > 400 nm), which can lead to cracking, severely impacting both device performance and yield. Despite considerable progress in thin-film processing, the reliable fabrication of high-confinement, high-yield Si$_3$N$_4$ platforms while maintaining high performance —particularly on foundry-grade 8-inch or larger wafers —remains a significant challenge.

In this work, we address this obstacle using a modified subtractive fabrication process along with optimized crack-isolation trench design, demonstrating reliable fabrication of 8-inch, crack-free, dispersion-engineered Si$_3$N$_4$-SiO$_2$-Si wafers utilizing a standard industrial foundry fabrication line. By combining an amorphous silicon (a-Si) hardmask etching technique, we achieve robust fabrication of ultra-low-loss Si$_3$N$_4$ PICs with intrinsic quality factors (Q$_i$) exceeding $25 \times 10^6$ using e-beam lithography (EBL) and $24 \times 10^6$ using ultraviolet (UV) stepper photolithography. To verify the designed anomalous dispersion, we demonstrate the generation of frequency combs. This work marks a significant step toward foundry-grade fabrication of scalable, high-performance, dispersion-engineered Si$_3$N$_4$ PICs for advanced nonlinear and quantum photonic applications.

\section{Results}

\begin{figure}[ht]
\centering
\includegraphics[width=\linewidth]{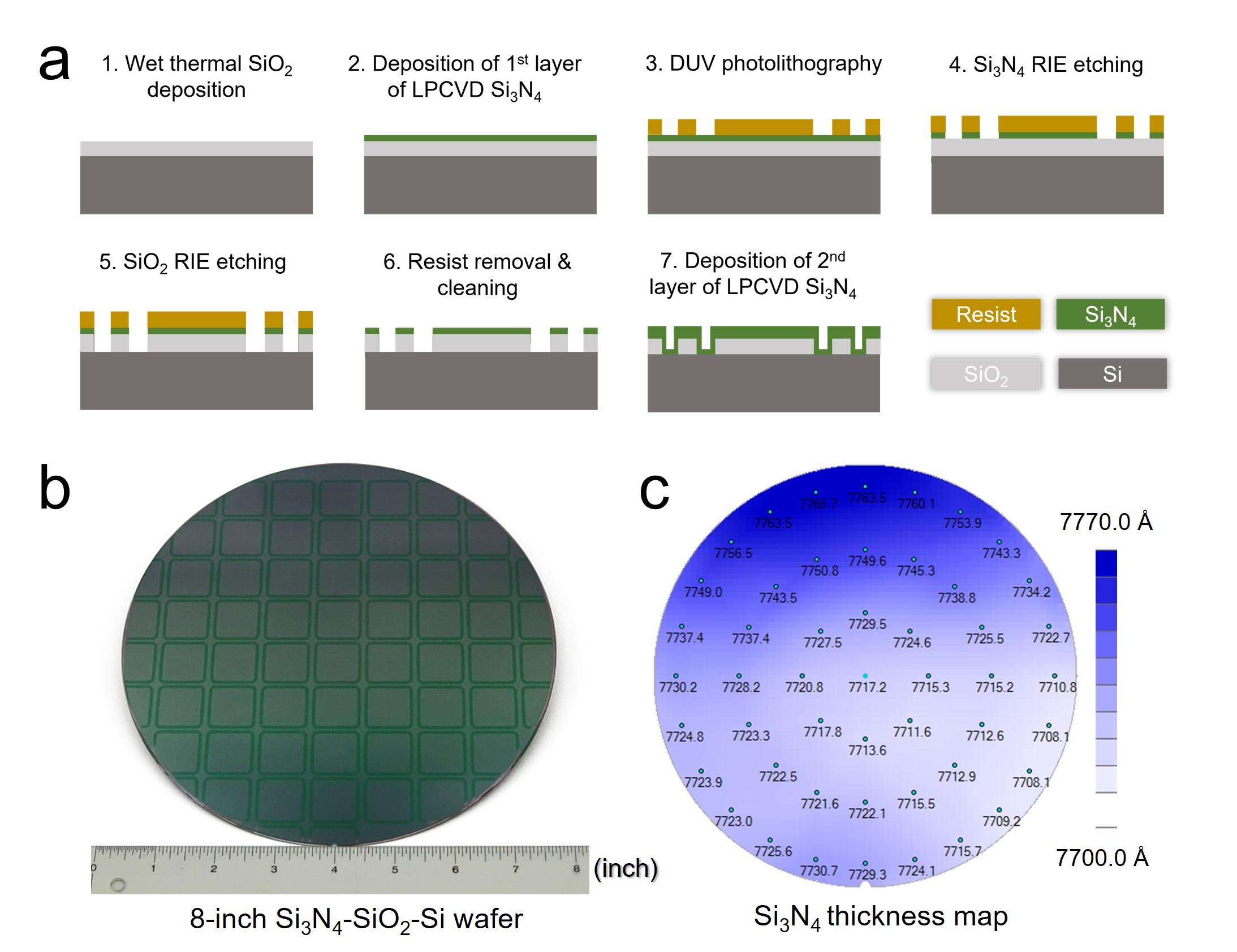}
\caption{\textbf{8-inch crack-free, anomalous-dispersion Si$_3$N$_4$-SiO$_2$-Si wafer:} 
    (a) Schematic illustration of the fabrication flow. (b) Photograph of the fabricated 8-inch Si$_3$N$_4$-SiO$_2$-Si wafer. (c) Thickness mapping of the Si$_3$N$_4$ film, showing a uniformity of $\pm 0.4$\% and mean thickness of 772.9 nm.}
    \vspace{-10pt} 
\end{figure}

The wafer fabrication is conducted at Honeywell's 8-inch silicon photonics fabrication facility \cite{puckett2021422}. To fabricate crack-free, dispersion-engineered 8-inch Si$_3$N$_4$-on-SiO$_2$-Si wafers, we adopt our previously developed subtractive processing approach and designs of cracking-isolation trenches \cite{liu2024manufacturing, liu2024fabrication}. As illustrated in Fig. 1(a), the process begins with standard 4-$\mu$m wet thermal oxidation and the deposition of a thin ($\sim$380 nm) low-pressure chemical vapor deposition (LPCVD) Si$_3$N$_4$ layer. Deep-ultraviolet (DUV) photolithography and  inductively coupled plasma reactive ion etching (ICP-RIE) are then employed to etch through the Si$_3$N$_4$ and SiO$_2$ layers to define the crack-isolation trenches along the boundaries of each exposure die. This trench design effectively suppresses crack propagation while preserving a large patternable area \cite{liu2024fabrication}. After resist removal and thorough wafer cleaning, a second LPCVD Si$_3$N$_4$ layer is deposited to achieve the target thickness of 775 nm. Figure 1(b) shows a photograph of the fabricated 8-inch Si$_3$N$_4$-on-SiO$_2$-Si wafer. The proposed fabrication strategy and cracking-isolation designs demonstrate exceptional robustness. Inspections of each single die on two fabricated 8-inch wafers reveal no cracking within all the protected regions, even after a long-haul transportation from Honeywell’s cleanroom to University of Michigan. Furthermore, thickness mapping of the wafers indicates an average Si$_3$N$_4$ film thickness of 772.9 nm with a uniformity of $\pm 0.4$\% as shown in Fig. 1(c), enabling the following precise dispersion control across the entire wafer.

\begin{figure}[b!]
\centering\includegraphics[width=\linewidth]{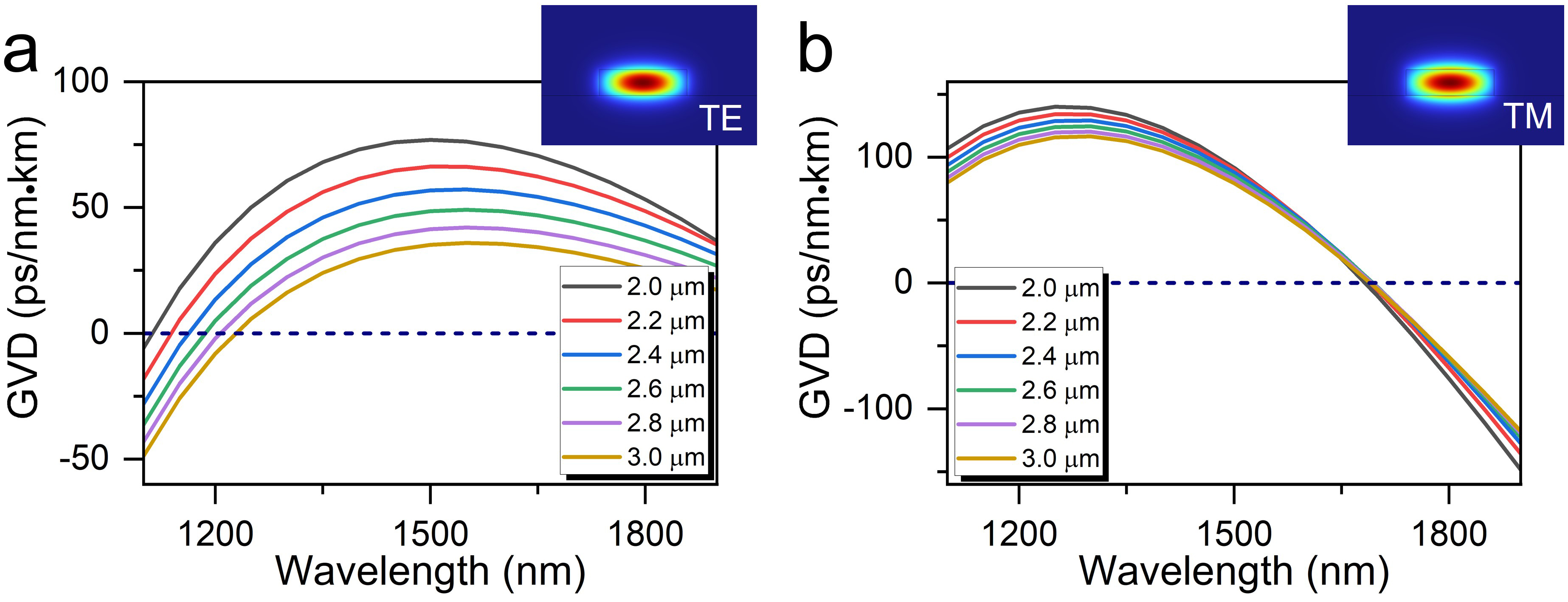}
\caption{Numerically simulated group velocity dispersion (GVD) of the Si$_3$N$_4$ waveguide with a fixed thickness at 773 nm and varying widths for the TE$_{00}$ (a) and TM$_{00}$ (b) modes respectively, demonstrating anomalous dispersion (GVD $> 0$) across a wide wavelength range.}
\end{figure}

\begin{figure}[ht!]
\centering\includegraphics[width=7.5cm]{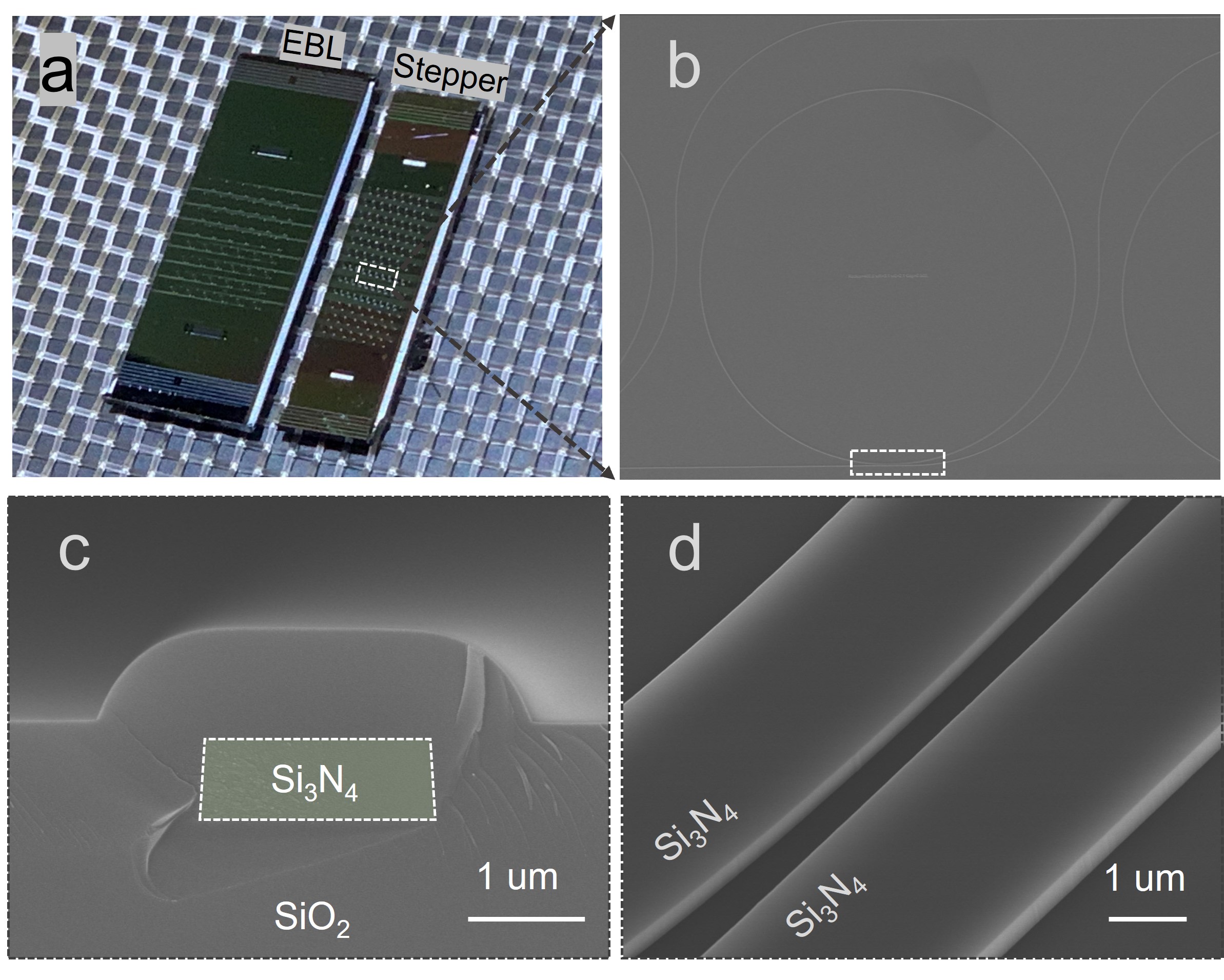}
\caption{\textbf{ Fabrication of Si$_3$N$_4$ photonic chips:}
(a) Photograph of the Si$_3$N$_4$ photonic chips fabricated using EBL and UV stepper photolithography.
(b)-(d) SEM images of the fabricated microring resonator, cross-section view, and tilted view of the Si$_3$N$_4$ waveguide, showing smooth boundaries that yield high $Q$ factors.}
\end{figure}

Figure 2 shows the numerically simulated group velocity dispersion (GVD) of a straight Si$_3$N$_4$ waveguide for the fundamental TE$_{00}$ and TM$_{00}$ modes, calculated using the commercial software COMSOL mode analysis. With the Si$_3$N$_4$ film thickness of 773 nm, anomalous dispersion (GVD > 0) is consistently achieved across a wide range of waveguide widths near 1550 nm wavelength band. These multimode waveguide dimensions not only provide flexible dispersion control but also reduce the interaction with sidewall roughness, contributing to lower propagation loss.

The subsequent fabrication of Si$_3$N$_4$ PICs is carried out at the University of Michigan using our recently developed a-Si hardmask etching technique \cite{liu2024fabrication}. This method combines a high etch rate, excellent etching selectivity, smooth and vertical waveguide sidewalls, and robustness to variations in the RIE chamber environment. Figure 3(a) shows a photograph of the fabricated Si$_3$N$_4$ PICs using EBL and UV stepper photolithography. For EBL exposure, the samples are patterned using a JEOL 6300 system with maN 2405 resist, supporting fabricaiton of small features. For UV stepper exposure, the GCA AS200 system with SPR 955 photoresist is employed, which aligns well with the foundry-grade DUV stepper photolithography for mass production. For both samples, the resist patterns are subsequently transferred into the a-Si hardmask layer via ICP-RIE etching using the LAM 9400 system with HBr and He gases. The Si$_3$N$_4$ layer is then etched using another ICP-RIE system (STS APS DGRIE) with a gas mixture of C$_4$F$_8$, CF$_4$, and He. The residual a-Si hardmask is removed using XeF$_2$ etching (Xactix), followed by RCA cleaning and high-temperature annealing at 1100°C to drive out the hydrogen. Finally, a SiO$_2$ cladding layer is deposited and annealed before testing. Figure 3(b) presents an scanning electron microscope (SEM) image of the fabricated Si$_3$N$_4$ microring resonator. The vertical sidewalls of the bus waveguide are clearly visible in the cross-section false-color SEM image shown in Fig. 3(c). And a zoomed-in view of the bus-to-resonator gap region is depicted in Fig. 3(d). No obvious sidewall and surface roughness is observed, indicating the high quality with this fabrication process.

\begin{figure}[ht!]
\centering\includegraphics[width=\linewidth]{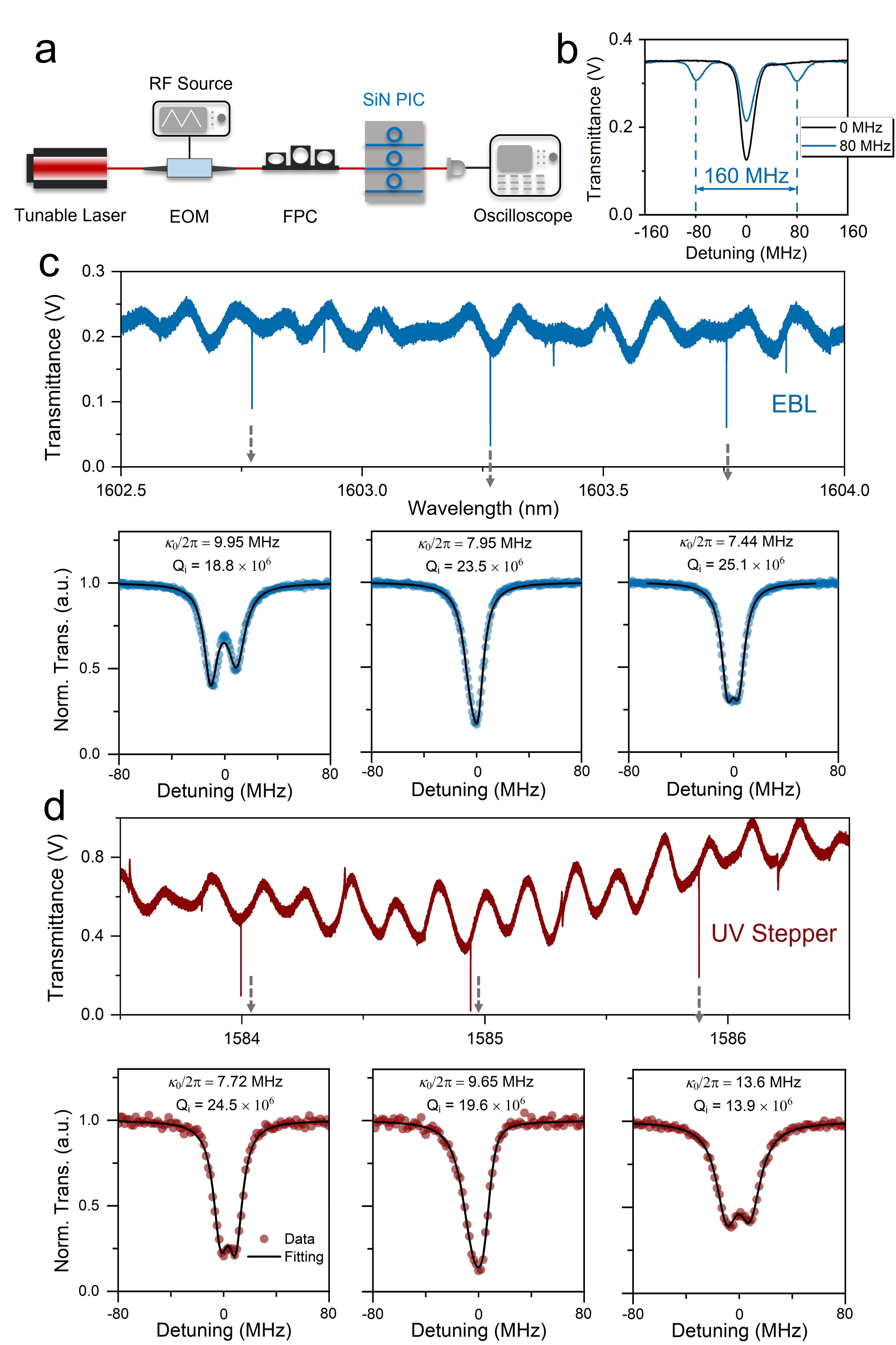}
\caption{{\textbf{Characterization of the Si$_3$N$_4$ photonic chips:}} (a) Schematic of the experimental setup used for resonance measurement and calibration with RF modulation markers. EOM: electro-optic modulator; FPC: fiber polarization controller; (b) Resonance spectra with and without 80 MHz RF-modulated sidebands. (c) Measured transmission spectrum near 1603 nm of a microring resonator fabricated by EBL with a radius of 400 $\mu$m and a waveguide width of 2.8 $\mu$m, demonstrating $Q_i \sim 25.1 \times 10^6$. (d) Measured transmission spectrum near 1585 nm of a microring resonator fabricated by UV stepper with a radius of 200 $\mu$m and a waveguide width of 2.7 $\mu$m, demonstrating $Q_i \sim 24.5 \times 10^6$.}
\end{figure}

To characterize the performance of the fabricated Si$_3$N$_4$ PICs, we measure the transmission spectra of microring resonators. The schematic of the experimental setup is shown in Fig.~4(a). A continuous-wave (c.w.) tunable laser (Santec TSL770) is used to sweep the laser wavelength across the resonances. To minimize wavelength uncertainty during the sweep, we employ a fiber electro-optic modulator (EOM) with a radio frequency (RF) signal applied to generate two sideband signals as calibration markers before injection onto the Si$_3$N$_4$ PICs. These markers provide precise calibration of the x-axis of the transmission spectra near each targeted resonance. To prevent thermal effects, the input laser power is limited to approximately 30 $\mu$W. Figure 4(b) presents an example of the measured transmission spectra with and without the RF markers. The two resonance dips adjacent to the central resonance correspond to the 80 MHz RF signal applied to the EOM, serving as reference points for wavelength calibration. 

Figure 4(c) shows the typical transmission spectra of transverse electric (TE) modes of a microring resonator fabricated using EBL, with a radius of 400 $\mu$m and a waveguide width of 2.8 $\mu$m. The bus waveguide width is set to 2.0 $\mu$m to ensure good phase matching with the fundamental TE$_{00}$ mode of the microring resonator. Using the 80 MHz RF markers, we fit three TE$_{00}$ resonances following the fitting strategy described in Ref. \cite{pfeiffer2018ultra}. For each resonance, both the intrinsic linewidth and coupling linewidth are extracted, while maintaining a coefficient of determination $R^2$ greater than 0.99. Across all three resonances, the measured data and the corresponding fitting curves exhibit strong agreement. The narrowest fitted intrinsic linewidth is $\kappa_0/2\pi$ = 7.44 MHz, corresponding to an intrinsic quality factor $Q_i \sim 25.1 \times 10^6$ and a propagation loss of 1.4 dB/m \cite{luke2013overcoming}. 

Figure 4(d) presents the transmission spectra for transverse magnetic (TM) modes of a microring resonator fabricated using UV stepper photolithography, with a radius of $200$ $\mu$m and a waveguide width of $2.7$ $\mu$m. The highest measured intrinsic quality factor near 1584 nm is $Q_i \sim 24.5 \times 10^6$, corresponding to a propagation los of 1.5 dB/m. This high-\textit{Q} factor is attributed to the robustness and stability of the a-Si hardmask etching technique, with performance comparable to EBL. The resolution limit of our UV stepper system is approximately 500 nm. Additional improvements are anticipated using a foundry-grade DUV stepper system along with the developed a-Si hardmask etching technique, which could potentially offer the combined benefits of ultralow optical loss, mass production, and the ability to define small feature sizes, e.g. tapered waveguide and grating coupler.

\begin{figure}[th!]
\centering\includegraphics[width=\linewidth]{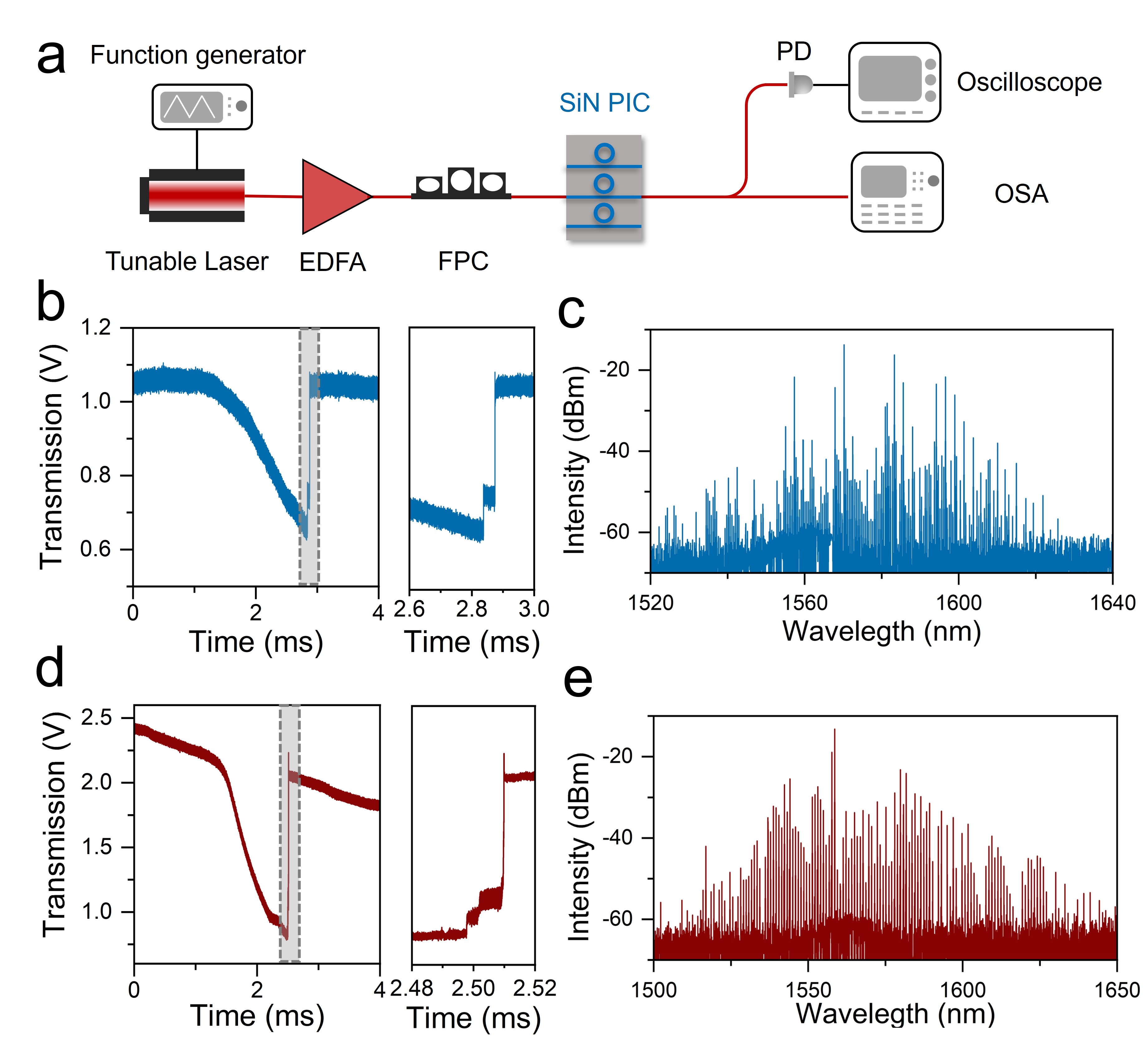}
\caption{\textbf{Frequency-comb generation in Si$_3$N$_4$ photonic chips:}
(a) Experimental setup for frequency-comb generation. EDFA: erbium-doped fiber amplifier; OSA: optical spectrum analyzer; PD: photodetector; FPC: fiber polarization controller.
(b) Typical transmission spectrum of the EBL-fabricated microring resonator at an input power of about 100 mW, with a zoomed-in view highlighting the soliton step.
(c) Frequency-comb spectrum of the chaotic modulation instability state.
(d) and (e) show the transmission spectrum and the generated frequency combs of the microring resonator fabricated using UV stepper lithography.}
\end{figure}

The generation of broad Kerr frequency combs, driven by third-order nonlinearity and tailored anomalous dispersion, has made thick Si$_3$N$_4$ PIC platforms highly effective for various important applications in nonlinear optics and quantum optics. To demonstrate frequency-comb generation using the fabricated Si$_3$N$_4$ PICs, we employ the experimental setup illustrated in Fig. 5(a). A c.w. laser is amplified by an erbium-doped fiber amplifier (EDFA) to ensure an on-chip pump power of approximately 100 mW. An external function generator is used to tune the pump wavelength with controllable sweeping speed across the targeted resonances. The output signal is then collected and split into two paths: one arm is directed to a photodetector (PD) and monitored in real time using an oscilloscope to track the transmission power; the other arm is sent to an optical spectrum analyzer (OSA) to display the generated Kerr comb spectra. Figure 5(b) presents a typical resonance scan from blue to red detuning of the resonator used in Fig. 4(c), with a zoomed-in view highlighting the clear discrete soliton step on the effectively red-detuned side. The corresponding chaotic modulation instability (MI) frequency comb spectrum is shown in Fig. 5(c). Additional transmission spectra, soliton steps, and MI frequency comb spectra are also provided in Fig. 5(d) and 5(e), corresponding to the stepper-fabricated sample of Fig. 4(d). Several techniques can be employed to access soliton states, including power kicking \cite{yi2016active}, rapid laser scanning \cite{herr2014temporal}, and bidirectional pumping for thermal compensation \cite{zhou2019soliton}. Overall, our experimental results highlight the potential of foundry-level fabrication of high-performance Si$_3$N$_4$ PICs for advanced applications in nonlinear optics and quantum information processing.

\section{Conclusion}

In summary, we have demonstrated the reliable fabrication of 8-inch, crack-free Si$_3$N$_4$-SiO$_2$-Si wafers with a Si$_3$N$_4$ thickness exceeding 775 nm. This platform enables the realization of anomalous dispersion in Si$_3$N$_4$ waveguides, a critical requirement for nonlinear optics applications. The fabricated ultra-low-loss Si$_3$N$_4$ PICs exhibit $Q_i$ exceeding $25 \times 10^6$ and are fully compatible with standard silicon photonics foundry processes. This work addresses a long-standing challenge in the mass production of high-confinement Si$_3$N$_4$ waveguides, particularly on 8-inch or larger-level foundry-compatible wafers, paving the way for high-performance, high-yield, and cost-effective manufacturing of dispersion-engineered Si$_3$N$_4$ PICs.

\begin{backmatter}
\bmsection{Funding} This work was supported by the National Science Foundation under Grant No.~2326780, No.~2330310, and No.~2317471 and University of Michigan.


\bmsection{Disclosures} The authors declare no conflicts of interest.

\bmsection{Data availability} Data underlying the results presented in this paper are not publicly available at this time but may be obtained from the authors upon reasonable request.

\end{backmatter}

\bibliography{sample}

\bibliographyfullrefs{sample}


\ifthenelse{\equal{\journalref}{aop}}{%
\section*{Author Biographies}
\begingroup
\setlength\intextsep{0pt}
\begin{minipage}[t][6.3cm][t]{1.0\textwidth} 
  \begin{wrapfigure}{L}{0.25\textwidth}
    \includegraphics[width=0.25\textwidth]{john_smith.eps}
  \end{wrapfigure}
  \noindent
  {\bfseries John Smith} received his BSc (Mathematics) in 2000 from The University of Maryland. His research interests include lasers and optics.
\end{minipage}
\begin{minipage}{1.0\textwidth}
  \begin{wrapfigure}{L}{0.25\textwidth}
    \includegraphics[width=0.25\textwidth]{alice_smith.eps}
  \end{wrapfigure}
  \noindent
  {\bfseries Alice Smith} also received her BSc (Mathematics) in 2000 from The University of Maryland. Her research interests also include lasers and optics.
\end{minipage}
\endgroup
}{}

\end{document}